\documentclass[11pt]{article}
\usepackage{acl2016}
\usepackage{times}
\usepackage{url}
\usepackage{latexsym}
\usepackage{amssymb}

\aclfinalcopy 

\title{Similarity in Observable Behaviors: A Synthesis of Studies with Implications for Socially-Aware Educational Technology Design \newline \newline \small{LTI Student Research Symposium 2016, Carnegie Mellon University, Pittsburgh PA 15213, USA}}

\author{Tanmay Sinha\\
  Language Technologies Institute\\School of Computer Science\\
  Carnegie Mellon University\\
  {\tt tanmays@cs.cmu.edu} }

\date{}

\begin{document}
\maketitle
\begin{abstract}
\vspace{-0.1cm}
Conversation is like an intricate partner dance and behavioral convergence, or the similarity in observable behaviors of partners over time, can lead to shared understanding, changed beliefs and increased rapport. This article describes a synthesis of three strands of our work on fine-grained analysis of conversational interaction in peer tutoring at the paralinguistic and verbal levels, in an attempt to better understand the phenomenon of behavioral convergence and its relationship to social and cognitive constructs. Implications for development of socially-aware agents that can improve task performance through convergence to and from the human learner's behavior are discussed.

\end{abstract}

\section{Introduction and Motivation}
\vspace{-0.1cm}
Similarity in observable behaviors or behavioral convergence, has been identified as one, largely subconscious contributor to successful conversations. In past theoretical and empirical work, this phenomenon has not only been shown to lead to communicative success of the interaction by decreasing misunderstandings and attaining goals faster \cite{18}, but also to its social success by building interpersonal closeness, engagement and affiliation \cite{14,55}, and task success by improving learning \cite{11}. 

It has been posited that similarity in verbal and non-verbal behaviors of partners engaged in an interaction can arise from a variety of sources, ranging from a tendency to focus on a comparison with others in new social environments \cite{festinger1954theory}, to a desire for approval and more efficient communication by reduction of communicative differences \cite{12}, to a consistent kind of non-conscious mimicry \cite{16}. 

Our work has taken multiple perspectives to examine behavioral convergence in the context of peer tutoring, a natural context for dyadic social interaction that provides opportunities for developing social and communication skills, positive attitudes towards social relationships and the process of learning, as well as domain-specific skills \cite{22}. Specifically, our study context comprises reciprocal peer tutoring data from 12 American English-speaking dyads (6 friends and 6 strangers; 6 boys and 6 girls), with a mean age of 13 years, who interacted for 5 hourly sessions over as many weeks (a total of 60 sessions, and 5400 minutes of data), tutoring one another in algebra. Each session began with a period of getting to know one another, after which the first tutoring period started, followed by another small social interlude, a second tutoring period with role reversal between the tutor and tutee, and then the final social time. 

The {\bf first strand of work (Study 1)} looked at automatically harvestable para-linguistic cues such as speech rate, laughter, overlaps etc for each speaker in the conversation for every 30 second, and operationalized the degree to which the speakers became similar along these behavioral dimensions \cite{sinha2015fine,sinhaalign}. The {\bf second strand of work (Study 2)} looked at reliably annotated relationship-oriented conversational strategies (particular ways of talking) that conversational partners use to regulate their interaction, and operationalized similarity in the cumulative as well as temporal usage of these verbal behaviors \cite{sinhastrategy}. Formalisms from time-series analysis \cite{kruskal1983symmetric,bernard1996interpreting} were used to examine the mutuality and dynamics in observable behaviors, accounting for the interdependencies between speakers in the longitudinal interaction. The {\bf third strand of work (Study 3)} moved beyond looking at just communicative behaviors to peek into the mental models, or the internal representations of concepts and the relationships \cite{novak2006theory}, for the speakers engaged in the interaction, and operationalized the sharedness in partner's mental models from their dialog transcripts. Formalisms from network text analysis \cite{carley1997extracting} were used to computationally generate and intersect the concept maps to look at structural aspects of the shared mental models.
\vspace{-0.1cm}

\section{Rationale}
\vspace{-0.2cm}
We followed foundational work that studies language as a form of joint action \cite{51}, challenging the traditionally held assumption in cognitive psychology of perception, and higher-level cognitive processes being best understood by investigating individual minds in isolation. Accordingly, the dyad was the unit of analysis. In addition, our work also aligned with the psycho-linguistic contribution theory \cite{clark1991grounding}, in the sense that we analyzed the individual contributions as expressions of their mental representations or personal beliefs, and treated the resultant shared understanding as a matter of similar mental contents or acceptance of pre-conceived beliefs in time. 

In other words, we followed the rationalist paradigm of shared understanding \cite{stahl2015essays}, which assumes individuals each have a stock of propositions in their minds that represent their current beliefs and opinions and the notion of shared understanding starts from individual understanding of two people and tries to establish equivalence of one or more propositions they hold. In the collaborative learning literature, this has often been referred to as cognitive convergence \cite{teasley2008cognitive} or convergence of two mental models (sharing as mutual giving), and in the group science literature, this has been referred to as team mental models \cite{klimoski1994team}.

Since this research is part of a larger research program on the social infrastructure of learning with an eye towards implementing more socially-aware educational technologies\footnote{http://articulab.hcii.cs.cmu.edu/projects/rapt/}, our studies leveraged concrete task outcomes such as pre-/post-test learning gains (paired t-test: t = 1.6964, df = 21, p $<0.1$) and problem-solving performance (total number of problems solved/attempted), as well as concrete social outcomes such as interpersonal rapport (the feeling of harmony or connection with another), in order to assess the impact of similarity in observable behaviors. Rapport was measured in two ways: a)self report via questionnaire, where rapport was broken into its component parts of attentiveness (Cronbach $\alpha=$0.42) , positivity (Cronbach $\alpha=$0.72) and coordination (Cronbach $\alpha=$0.64) \cite{25}, and b)perceived rapport via annotation by naive raters of every 30 second thin-slice \cite{1} of the video of the peer tutoring on a scale of one to seven, presented to the annotators in a randomized order (Intra-Class Correlation=0.37, Cronbach $\alpha=$0.68)

Specifically, the kind of educational technologies we are interested in building are a new genre of intelligent tutoring systems - virtual peer tutors who can sigh in frustrated solidarity about a learning task at hand, assess the state of a relationship with a student and know how to respond to maximize learning in the peer tutoring context. Such virtual peers, whose underlying behavioral models are derived from close observation of human-human interaction, may particularly encourage learning among populations not served well by current educational practices.  In her seminal work on embodied conversational agents, \cite{cassell1999requirements} emphasizes ``multi-threaded entrainment", meaning adapting towards or synchronization of visual, vocal and verbal modalities between the speaker and interlocutor, as one of the primary design drivers for computational systems that can support real-time multimodal interaction. In that vein, the understanding of different channels along which such behavioral similarity manifests in human-human tutoring interaction, is essential to plug similar social capabilities into pedagogical agents that increasingly participate in human-agent interactions \cite{johnson2016face}, as a path towards increasing adaptivity and naturalness in the social interaction. 
\vspace{-0.15cm}

\section{Study 1: Paralinguistic Convergence}
\vspace{-0.2cm}
\subsection{Method}
\vspace{-0.2cm}
{\bf Step 1:} In order to study similarity in paralinguistic behaviors, we computed the following features for each speaker for each consecutive 30 second segment in the tutoring sessions: a) number of words spoken, b) message density, which is the number of independent clauses uttered, divided by the time difference between the first and last utterance within the 30 second segment, c) content density, which is the number of characters spoken divided by the number of independent clauses uttered, d) number of overlaps (a joint event where the two interlocutors speak at once), and e) number of laughter expressions. The first three features are representative of speaking rate and two dimensions of burstiness that characterize it. Frequent turn taking will increase the message density, more elaborate or detailed talk between interlocutors will increase the content density, while more number of overlaps might potentially indicate a well-coordinated conversation. Laughter may signal enjoyment and affiliation, but also serves to release tension.

{\bf Step 2:} Next, we followed the tests of convergence hypothesis \cite{bernard1996interpreting} to quantify behavioral convergence. This involved: a) computing the difference in raw behavioral feature values for partner $i$ and partner $j$ engaged in the dyadic conversation for every 30 second slice (call this differenced series $y$), b) formulating the autoregressive model as $\Delta y_t =\alpha + \beta t+ \gamma y_{t-1}+ \delta_1 \Delta y_{t-1}+ ...+\delta_{p-1} \Delta y_{t-p+1}+\varepsilon_t$, where $\alpha$ (constant term) is the drift or change of the average value of the stochastic process, $\beta t$ is the deterministic time trend and $p$ is the lag length (which was quantified as 3, meaning we looked back at the time window of 90 seconds (30*3), i.e  time slices `t-1', `t-2' and `t-3' when predicting the similarity in behaviors at time slice `t'), c) testing the presence of unit-root in this time series framework using the Augmented Dickey Fuller (ADF) test at 1\% level of significance. 

Intuitively, if the ADF test statistic was significant, we rejected the null hypothesis that the differenced behavioral time series had a unit root and accepted the alternative hypothesis that the variable was generated by a stationary process, which was an evidence for convergence. On the contrary, if ADF test statistic was not significant, we accepted the null hypothesis of the presence of a unit root, in turn indicating that the process (change) was not stationary and the definition of convergence was violated. 

{\bf Step 3:} To construct a composite score for paralinguistic convergence or the convergence strength, the ADF test statistic (call this $x$) along each feature dimension was firstly scaled between 0 and 1 using the formula $(x-minimum(x))/(maximum(x)-minimum(x))$, with an intuition to provide transparency and comparability. Secondly, in weighting across features, different feature dimensions were equally weighted (averaged).
\vspace{-0.15cm}

\subsection{Results}
\vspace{-0.2cm}
Preliminary results revealed that virtually all dyads and all sessions seemed to converge on message density (roughly, speech rate). This bi-directional accommodation is the kind of classic entrainment that indexes engagement in the interaction - engagement that we also saw indexed with the content of the conversations.

On correlating our measures of convergence strength with learning, we found positive correlation (Pearson $r$=$0.57$, $p$=0.06) between the maximum convergence strength of a dyad across different sessions and their composite learning gains (average of the individual learning gains). On further looking at the association between convergence strength and self-reported rapport, we found that when the dyad rated the attentiveness questionnaire dimension identically, their convergence strength was in fact lesser (Point Biserial correlation $r_{pb}$=$-0.57$, $p<$0.05). This particular result echoes with the classic explanation of convergence operating as a non-conscious phenomenon.

Finally, we performed a causal analysis to investigate whether rapport led to convergence in the longitudinal interaction. This involved: a) formalizing the convergence time series (say $T_1$) as the difference in detrended and smoothed low level feature value (e.g, \#words spoken) for the two students in every dyad at lag 0, for every 30 second segment in the hourly session, b) formalizing the rapport time series (say $T_2$) as the detrended and smoothed value of thin slice rapport, c) inferring whether $T_2$ significantly granger-causes $T_1$ (at 5\% level of significance), using a time window of 90 seconds in the autoregressive model formulation, meaning that we additionally looked at the rapport rating for time slices `i-1', `i-2' and `i-3' in inferring about convergence at time slice `i' apart from just using convergence information from time 0 to `i-1'). Technically, the causality would be significant (assessed via an F-test), if the inclusion of past observations (lagged values) of $T_2$ reduced the prediction error of $T_1$ in a linear regression model of $T_1$ and $T_2$, as compared to a model including only the previous observations of $T_1$ \cite{13}. The F-statistic was taken as granger causality magnitude.

We found that for $\sim$87\% of the sessions used in our study, rapport significantly led to convergence in message density (that indexes speech rate in the interaction). To evaluate the robustness of our findings, we also altered the definition of convergence time series $T_1$ as the difference in detrended and smoothed low level feature value for the two students in every dyad at lag 1 and 2 respectively, the intuition being to include a time contingency effect for similarity in behaviors over time. However, the causal effects for rapport on convergence in message density were still significantly present for $\sim$87\% dyadic sessions. The second strongest causal effects were found for number of words ($\sim$60\% sessions) and number of overlaps ($\sim$53\% sessions), meaning that increased rapport led to similarity in \#words spoken and overlapping expressions used. Interestingly, by reversing the direction of causality and testing whether convergence and rapport work together in a feedback loop (rapport $\Leftrightarrow$ convergence) to help in interaction regulation, we found very low support (roughly $1/3^{rd}$ sessions) for significant causal effects of convergence on rapport in the dyadic sessions, along any of the feature dimensions. 
\vspace{-0.15cm}

\subsection{Mechanism and Implications}
\vspace{-0.2cm}
In summary, we found that: a) Paralinguistic convergence had a positive effect on learning - a virtual peer tutor that both converges to or mimics its human partner and invites convergence by producing salient and highly mimicable behaviors may be a more effective learning partner, b) There was a significant causal effect of rapport on convergence - a virtual peer that builds rapport might lead to students speaking and behaving like the virtual peer. 

In order to explain these results, one might turn to the following mechanism. Learning can be explained as the side effect of cognitive \cite{dillenbourg2007designing} and social \cite{kreijns2003identifying} processes triggered by the conversational interactions (explanation, argumentation, mutual regulation etc) students engage in so as to develop shared understanding. The collaborative learning literature \cite{schwartz1995emergence} suggests that the effort necessary to build shared understanding \cite{teasley2008cognitive} is what actually leads to learning. In parallel, the entrainment literature provides theoretical evidence \cite{18} of convergence (alignment processes) being one of the important indices of shared understanding between interlocutors, allowing them to sufficiently reconstruct the meaning of the interaction. In their comprehensive synthesis of joint action studies, \cite{21} have emphasized that assessing the mutual influences of two or more actors on each other is an important step toward investigating the mechanisms whereby individuals coordinate their actions. 

Moreover, the Interactive Alignment Model \cite{18} also posits that such shared mental representation is caused by greater similarity in observable low-level behaviors (for e.g - lexical, acoustic, prosodic levels) and subsequently in internal (higher level semantic) representations. Interactive priming \cite{2b6bda00f2304921b2e0db0528df0a7b}, which links these neighboring levels of representation, has been described as one of the underlying mechanisms for observable convergence. Empirically \cite{reitter2006computational} too, priming has been utilized for operationalizing convergence in the dialog as a visible measure of shared mental models, which was in turn shown to be positively associated with learning \cite{11,ward2007dialog}. 

Thus, in the results described above, we posit that when there is rapport, it leads to convergence in the interaction - this social phenomenon of convergence causes shared mental representation, which in turn leads to learning. This has direct implications for the development of virtual companions that can improve students' performance by establishing positive relationship with them and in turn also potentially triggering verbal and non-verbal behavioral convergence. Such a form of social facilitation and socially adaptable behavior generation by a virtual agent has been shown to have a positive effect on student effort and performance on math tasks \cite{15}. 
\vspace{-0.2cm}

\section{Study 2: Conversational Strategy Convergence}
\vspace{-0.2cm}
\subsection{Method}
\vspace{-0.2cm}
{\bf Step 1:} In order to study similarity in usage of relationship-oriented communicative strategies used by interlocutors, we drew on our former work that has developed a dyadic computational model \cite{25,submitted1} to explain how dyads manage rapport through the use of specific conversational strategies \cite{submitted2}, which in turn function to fulfill social goals that make up rapport - face management, mutual attention, and coordination \cite{spencer2008face}. This served as a base for us to select which conversational strategies to examine. Face-boosting strategies such as {\em praise} serve to create increased self-esteem in the individual and increased interpersonal cohesiveness or rapport in the dyad. Mutual attention facilitates learning about the other person by leading dyads to provide information about themselves through the strategy of {\em self-disclosure}. In addition, over time interlocutors increasingly coordinate less to sociocultural norms set by the outside world, and more to interpersonally determined norms and {\em referring to shared experience} allows them to index commonality and differentiate in-group and out-group norms.

{\bf Step 2:} After making this initial selection, we employed 3-5 human annotators to code the three conversational strategies of self disclosure (Krippendorff's $\alpha$=0.753), reference to shared experience (Krippendorff's $\alpha$=0.798) and praise (Krippendorff's $\alpha$=1.0). After achieving high enough inter-rater reliability, most of the sessions were coded by independently by the annotators. Self-disclosure was defined as verbal expressions used by people to reveal aspects of themselves to others and comprised enduring states (that reveal the long-term aspects of oneself, that one may feel are deep and true, and therefore important to reveal in the context of a relationship. e.g, ``I love playing hockey") and transgressive (forbidden or socially-unacceptable) actions, which may be a way of making the other person feel better by disclosing that one is not perfect (e.g, ``I did badly on the pre-test"). 

Referring to shared experience is an important way of showing that the two interlocutors have known each other and interacted outside of the context of the current peer tutoring interaction, and comprised ROE (reference to outside current experience, meaning shared activities that are outside the experiment; e.g, ``Did you see my Facebook post last night?") and RIE (reference to inside current experience, meaning peer tutoring related experience. e.g, ``I remember you helped me with a problem like this before"). Praise is the expression of a favorable judgment of an attribute, behavior or product of the other person, and comprised labeled praise, or an expression of a positive evaluation of a specific attribute, behavior or product of the other person (e.g, ``great job with those negative numbers"), and unlabeled praise, which is a generic expression of positive evaluation, without a specific target (e.g, ``Perfect"). 

{\bf Step 3:} Finally, to quantify conversational strategy convergence, we utilized dynamic time warping (DTW) \cite{kruskal1983symmetric} to obtain a global distance that could characterize how conversational strategy usage for each partner in the dyad was aligned in time. Concretely, given these two time series, say $A = [a_1,...a_n] \in \mathbb{R}^{1Xn}$ and $B = [b_1,...b_n] \in \mathbb{R}^{1Xn}$, DTW is a technique to align $A$ and $B$ such that the sum of the Euclidean distances between the aligned samples is minimized. In order to perform this alignment, DTW can distort or warp the time axis, compressing it at some places and expanding it at others. In our case, each element in the two time series refers to time from the start of a peer tutoring session (in seconds) at which each individual in the dyad used certain conversational strategy. 

In terms of specific parameters employed, we chose: a) step-pattern (that lists transitions allowed while searching for the minimum distance path between the two time series), with the constraint that one diagonal step costs as much as the two equivalent steps along the sides, b) open-ended alignment, meaning that we freed the endpoint of time series $B$ in order to allow for a partial match, c) normalization of the DTW distance by the length of the two input time series, in order to accommodate time-series of varying lengths (certain dyads, for instance, do lot more self-disclosure relative to other dyads).
\vspace{-0.15cm}

\subsection{Results}
\vspace{-0.2cm}
We found that the less similar a dyad was in their pattern of timings of self-disclosure (lesser tendency for reciprocity), the more problems they solved (Spearman Rank $\rho$=$0.31$, $p$=0.031). For friend dyads, in addition, greater similarity in pattern of timings of self-disclosure was positively associated with higher ratings for attentiveness (Spearman Rank $\rho$=$-0.37$, $p$=0.04) and coordination (Spearman Rank $\rho$=$-0.37$, $p$=0.04) on the questionnaire dimension of self-reported rapport. These results suggest a positive impact of similarity in timings of conversational strategy usage by partners on rapport, while a negative impact on problem-solving. 

The patterns for reference to shared experience did not follow a similar trend. There were no significant associations with problem-solving performance. However, the more similar a female dyad was in their pattern of timings of shared experience usage, higher were the total positivity ratings (Pearson$r$=$-0.62$, $p$=0.02). With respect to praise, we found that the more similar a dyad was in their pattern of timings of praise, higher were the total attentiveness (Pearson $r$=$-0.58$, $p$=0.02) and coordination (Pearson $r$=$-0.52$, $p$=0.04) ratings. For strangers, in addition, we found greater similarity in the pattern of timings of praise to be associated with higher ratings for positivity (Spearman Rank $\rho$=$-0.85$, $p$=0.03). Since praise also serves to boost face of the interlocutor, the relationship to positivity (that indexes friendliness and warmth) makes sense. Female dyads also had the exchange patterns of praise over time positively associated with ratings for attentiveness (Pearson $r$=$-0.73$, $p$=0.01) and coordination (Pearson $r$=$-0.68$, $p$=0.03).
\vspace{-0.1cm}

\subsection{Mechanism and Implications}
\vspace{-0.2cm}
To understand the counter-intuitive result of the positive correlation between the similarity in the timing of conversational strategy usage with interpersonal rapport, but its negative correlation with problem-solving, we must first acknowledge that every peer tutoring session has a relational goal (manage-relationship) as well as a transactional or instrumental goal (tutor-math). Interactional goals such as these form one of the bases of rapport \cite{spencer2008face}. However, we believe that, given limited time, it is difficult for interlocutors to simultaneously achieve both goals. In addition, extreme attention paid to relational goals (via many exchanges of self-disclosure) can lead to reduced performance in the task at hand, in turn affecting future rapport. 

Conversational strategy exchange patterns, which are both a function of time as well as the relationship status of individuals, can be perceived as social bids with advantages and disadvantages. For instance, there is always risk and benefit involved during self-disclosures \cite{petronio2012boundaries}. The costs of disclosing are increased vulnerability and less privacy. The benefits are increased trust, rapport and reciprocation, which could outweigh the costs. The similarity in timings of their usage signals how similar response repertoires - a kind of social reciprocity - that promotes the maintenance of further social interactions, develops over time. For example, \cite{lewis1975beginning} points out that peers are thought to acquire similar behavioral repertoires during shared experiences, and therefore they are mutually attracted and seek each other out for further interaction. 

More broadly, the social exchange theory, which defines social behaviors as an exchange \cite{Hill:1982} posits that during the development of relationships, social exchange is regulated by a series of obligations \cite{Emerson:1976} - how we feel entitled to respond based on the behaviors we expect from others (for e.g, desire to be approved of), and that reciprocity is a very important social norm in the early stages of a relationship \cite{Altman:1973-1}. One possible underlying mechanism behind the differences in conversational strategy convergence patterns for friends and strangers, stems from the arousal reinforcement theory \cite{berlyne1967arousal}, which suggests that being in an unfamiliar context arouses the child to a non-optimal level for interaction or exploration, while presence of a familiar peer reduces the child's arousal to a more optimal level.
\vspace{-0.1cm}

\section{Study 3: Shared Mental Models}
\vspace{-0.2cm}
\subsection{Method}
\vspace{-0.1cm}
Though some consider mental models to be inherently unobservable, due to their dynamic and partial nature, other researchers consider mental models to be an emergent structure that comes into being as an individual articulates their ideas, and is therefore observable. Here, we used peer tutoring dialog transcripts to automatically generate concept maps \cite{novak2006theory} from the proximity between exact words (or, the generalization of those words) as a dynamic externalization \cite{carley1997extracting} of students' mental models. 

{\bf Step 1:} The first step was to decide what words to use in concept map creation. This involved standard preprocessing of the the dialog transcripts such as fixing of typos, removal of punctuations and other symbols etc. Then, noise words such as pronouns (he, she, it etc.), prepositions (on, but, till etc.), and noise verbs (is, am, was etc.) were filtered. Next, we constructed a delete list to further reduce the noise from commonly occurring words that had little conceptual value. We started with the default extensive delete list from \cite{carley1997extracting} and modified it by adding other frequently occurring words in the transcripts (yeah, okay, just etc.) to create our own modified delete list. 

As a final pre-processing step, we applied a generalization thesaurus to avoid concept misclassification due to variances in the wording between individual participants, or variations in verb tenses and pluralization of words that are, in fact, the same concept. For instance, we generalized each of the four main mathematical operations in our linear algebra domain. For any word with the stem ``add", such as ``adds, adding, added" we generalized those to the concept ``add". We followed a similar approach for ``subtract", ``divide", and ``multiply". Then, though each algebra problem may have used a different variable and number, each was an instance of the overarching concepts ``variable" and ``number". 

{\bf Step 2:} The second step involved creation of the concept maps for the tutor and tutee by: a) specifying unidirectional linkages to allow for forward search in the pre-processed transcript while capturing proximity, b) following a windowing-based approach, choosing a small window-size of 10 words as the farthest distance from one word to another for a possible connection, and finally c) specifying the stop unit as 10 sentences indicating boundaries of the moving window across the transcript. The intuition was to determine which concepts should have relationships placed between them, and how proximally distant two concepts could be from each other and still have a relationship. The nodes in these concept maps (weighted directed networks) represent words while an edge, or link, $E_{ij}$ represents the number of times concept $i$ was found to be linked to concept $j$, given the windowing and stop unit constraints. 

We encoded two fundamental properties of the relationship that could exist between concepts - directionality and weight. Weight not only tells about the level of usage but also the emphasis in the text given to a certain relationship by the speaker. Also, note that since our pre-processing steps filtered out some concepts and generalized other words into larger conceptual categories, we didn't simply place relationships between every adjacent word. The filtered transcript allows us to be more selective in creating linked relationships between concepts.

{\bf Step 3:} The third step involved intersecting the tutor and tutee's concept maps for each tutoring period to facilitate comparison of individual concept maps. After the intersection, simple network statistics such as the numbers of shared concepts, shared links and the mean betweenness centrality for the intersected network were computed. Shared concepts are concepts that appear in both Student A and Student B's concept maps. Shared links in the intersected concept maps refer to the minimum value of the number of times link $E_{ij}$ occurred in both Student A and Student B's concept maps. Note that because we used unidirectional linkages, if Student A has a link $E_{ij}$ but Student B has a link $E_{ij}$ in their concept maps respectively, those would not register as shared links. 

The mean betweenness centrality in a concept map will be high if many words act as connecting words, or fall on the shortest path between usages of other words. Accordingly, a higher value for this mean betweenness in the intersection network signals that the tutor and tutee have a high overlap in their usage of central connecting words. Examples of words with high betweenness include the obvious such as ``number", ``variable", and the operation words such as ``add", ``subtract", etc., but also other essential conceptual terms that serve as connecting words in this domain, such as ``sides", ``simplify", ``terms", ``distribute", and ``isolate". 
\vspace{-0.15cm}

\subsection{Results}
\vspace{-0.15cm}
We used linear mixed effect models to predict our outcome measures of problem solving performance (dependent measure). The random effects in our model comprised the grouping factors - dyad (12 dyads overall in the reciprocal peer tutoring sessions) and session (1 through 5). Also, relationship status (friends or strangers), gender (male or female), pretest score and interpersonal rapport were the covariates we controlled for. The fixed effects of interest (independent measures) in our model included shared concepts, shared links and mean betweenness centrality from the shared map between the tutee and the tutor. We scaled both the dependent and independent measures by converting them to z-scores. 

Preliminary results revealed that shared concept maps could explain up to 65\% of variation in tutee's problem-solving performance. On looking at estimates, we found that higher number of non-isolated shared concepts (those that were associated with the shared concept map structure) between the tutor's and tutee's dialog, were positively predictive tutee's problem-solving ($estimate$=$0.54$). Note that we divided shared concepts into isolated and non-isolated categories, the former referring to concepts that were used by both the tutor and tutee but not in the same contextual manner, while the latter referring to concepts that were used in the same way (associated with shared links). In addition, the tutor using more concepts from the solved working sheet (higher number of shared concepts between tutor's dialog and instructions from solved working sheet) was positively predictive of tutee's problem-solving ($estimate$=$0.11$), but the tutor repeating solved worksheet steps ``verbatim" (high number of shared links between tutor's dialog and instructions from solved working sheet) was negatively predictive ($estimate$=$-0.46$)
\vspace{-0.25cm}

\subsection{Mechanism and Implications}
\vspace{-0.2cm}
Prior work in the educational literature has suggested positive impacts of both cognitive convergence \cite{teasley2008cognitive} and interactional convergence on learning. Such alignment processes, as visible instantiations of shared mental models not only signal shared understanding between interlocutors but also explain processes that move a group towards problem solving goals by making their knowledge explicit and establishing the common ground between individuals. In conversational interactions between a tutor and tutee, the effort required to achieve such shared understanding by continually seeking evidence for grounding is what leads to increased learning. Thus, the goal behind better understanding of alignment at different levels is to produce actionable interventions that could increase sharedness of mental models to better stimulate learning.

Our results bear important implications for providing adaptive scaffolding to students working in groups in a classroom or interacting with learning technologies such as tutoring systems. To maximize tutee performance, our results suggest a need to incorporate linguistic alignment support into the design of group activities in a way that not only evokes from the tutee similar concepts as the tutor, but also reinforces a shared contextual usage of those concepts, since higher numbers of non-isolated shared concepts between tutor and tutee are predictive of increased correctness in tutee problem solving. For the design of dialog-based educational technologies, this could take the form of goals or policies that work towards increasing the number of shared conceptual links between the tutoring agent and tutee in a real-time interaction. 

Thus, by providing adaptive scaffolding, one might be able to facilitate increased alignment of students' mental models with the tutor in a way that gives students a) more opportunities to demonstrate knowledge about the subject matter and b) more points to engage in deeper reflection or reasoning about the content (why, how, what-if etc.). For instance, by planting contradictions and paradoxes, the tutor might motivate a mutual discussion, encouraging self-regulation and self-correction of mistakes, in turn leading the tutee to use task-related concepts in a manner closely aligned with the tutor. 

Another such instructional strategy that could help achieve greater sharedness in concept maps of tutor and tutee over time could be explicit requests for verification or summary of the current problem solving step, in order to assess the student's knowledge, encourage them to stay engaged and generate information and provide a space for repetition of their learning experience. In addition, a shared concept map visualization could be provided to a tutoring dyad at the end of their session, or to the teacher, to help stimulate reflection on the effectiveness of the interactions in the tutoring process. This could be useful for providing real-time feedback to students, than comparison to an expert model post-hoc.

Finally, our work, which uses a semi-automatic approach to dynamically extract shared concept maps from transcripts of multiple tutoring sessions, favors reproducibility, speed and scalability. It addresses some existing issues in creation and evaluation of concept maps \cite{ruiz1996problems}. Typically, students themselves create these maps by choosing concepts from a predetermined list or filling blanks in a given conceptual structure \cite{chang2007externalising}. However, these self-created approaches  are time-consuming to manually create and evaluate post-hoc. The meta-cognitive act of creating concept map representations may introduce bias from students' awareness of being assessed \cite{andersen1997group}. Furthermore, these maps may be inconsistent when created by the same student over time. 

As far as evaluation is concerned, the typical method is through comparison between an individual's concept map and an expert model for that domain or problem space \cite{harnisch1994concept}. However, in cases where there are no expert models readily available, a different method is required. Our work, then, employs a simple network analytic lens to describe the impact of convergence between tutor and tutee's dialog contributions on tutee performance over time. 
\vspace{-0.15cm}

\section{Conclusion}
\vspace{-0.2cm}
Dyads participating in peer tutoring require a pattern of cognitive similarity that enables the tutor and tutee to anticipate one another's needs and actions and synchronize their work in a way that is synergistic toward meeting the dyad's ultimate goals. One salient observable manifestation of such synchronization is behavioral convergence, which can happen at both the para-linguistic and verbal levels. This article described three studies that aimed to gain a better understanding of how individuals engaged in a dyadic interaction exhibited similarity in behaviors and whether and why this was associated with increased task performance and interpersonal closeness between them. Our goal is to have a roadmap for integrating convergence into our dialog-based reciprocal peer tutoring virtual agent, in such a way as to evoke alignment, and to detect and remedy decreasing alignment, between the tutor and student in real time. 
\vspace{-0.2cm}

\section*{Acknowledgments}
\vspace{-0.2cm}
Thanks to Justine Cassell for providing the opportunity to engage in this exciting research. Many thanks to Ran Zhao, Michael Madaio and other current and former ArticuLab members for wonderful discussions that influenced this work.

\vspace{-0.7cm}
\bibliography{acl2016}
\vspace{-0.7cm}
\bibliographystyle{acl2016}

\end{document}